\documentclass[10pt]{aiaa-tc} 

\usepackage{epsf,exscale,times}
\usepackage{varioref}
\usepackage{ams}
\usepackage{graphicx}

\usepackage{subcaption}
\usepackage{epsfig}
\usepackage{wrapfig}
\usepackage{lettrine}
\usepackage{amssymb}
\usepackage{amsthm}

\usepackage{booktabs}
\usepackage[sort&compress,numbers]{natbib}
\usepackage{setspace}

\usepackage{xcolor}

\usepackage{amsmath}

\usepackage{tikz}
 \usetikzlibrary{patterns}        
 \usetikzlibrary{arrows,automata}
\usetikzlibrary{calc}

\newsavebox{\astrutbox}
\sbox{\astrutbox}{\rule[-5pt]{0pt}{20pt}}

\usepackage{tikz}
\usepackage{verbatim}

\usetikzlibrary{arrows,positioning} 
\tikzset{
    >=stealth',
    punkt/.style={
           rectangle,
           rounded corners,
           draw=black, very thick,
           text width=6.5em,
           minimum height=2em,
           text centered},
    pil/.style={
           ->,
           thick,
           shorten <=2pt,
           shorten >=2pt,}
}

\tikzset{
   punkt0/.style={
           rectangle,
                      fill=yellow,
           rounded corners,
           draw=black, very thick,
           text width=6.5em,
           minimum height=2em,
           text centered},
    pil/.style={
           ->,
           thick,
           shorten <=2pt,
           shorten >=2pt,}
}

\tikzset{
    >=stealth',
    punkt2/.style={
           rectangle,
           rounded corners,
           draw=black, very thick,
           text width=10em,
           minimum height=2em},
    pil/.style={
           ->,
           thick,
           shorten <=2pt,
           shorten >=2pt,}
}

\tikzset{
    >=stealth',
    punkt3/.style={
           rectangle,
           rounded corners,
           draw=black, very thick,
           text width=10em,
           minimum height=2em},
    pil/.style={
           ->,
           thick,
           shorten <=2pt,
           shorten >=2pt,}
}

\tikzstyle{xline}=[myblue,thick]
\colorlet{myblue}{blue!80!black}

 \title{Statistical Turbulence and High-Fidelity Disturbance Fields for Quadrotor Flight Control} 
 
 \author{Xun Huang %
    \thanks{Professor. State Key Laboratory of Turbulence and Complex Systems, Department of Aeronautics and Astronautics.  huangxun@pku.edu.cn. AIAA Associate Fellow.    }
\\   
 {  \normalsize\itshape
    Peking University, Beijing, 100871, People's Republic of China}\\
 }

\date{}
\begin{document}

\maketitle

\begin{abstract}
Reinforcement-learning quadrotor controllers are usually trained under simplified wind models, yet the impact of wind-field fidelity---as opposed to magnitude---on policy robustness remains unquantified. This paper compares five disturbance-fidelity levels, from wind-free flight and discrete 1-cosine gusts through statistical turbulence and synthetic coherent structures to large-eddy-simulation fields of the atmospheric boundary layer, in a full cross-fidelity train--test evaluation of proximal policy optimization (PPO) agents, with cascaded PID and geometric SE(3) controllers as training-free references, over a 0--12~m/s wind sweep. Before any controller comparison is made, all disturbance data are validated: every synthetic generator is checked quantitatively against its analytical or certification-standard reference, and the large-eddy-simulation fields against the imposed log law. On a racing-class quadrotor in hover, the train$\times$test matrix is remarkably flat, and the cheapest structured training wind---discrete-gust domain randomization---ranks first in every test column, a ranking replicated on a wind-sensitive 27~g platform; a once-tuned geometric controller brackets the learned PPO policies at zero crash rate. Mechanism diagnostics show that control authority, not wind realism, bounds robustness, so wind-fidelity investment should scale with platform wind sensitivity.
\end{abstract}

\section*{Nomenclature}


 \begin{tabbing}
\hspace{4cm} \= \kill %
$\mathbf{w}(\mathbf{x},t)$ \> \hspace{2mm} = \hspace{2mm} three-dimensional wind velocity vector, m/s \\
$U(z), U_\infty$ \> \hspace{2mm} = \hspace{2mm} mean wind speed profile; reference mean speed, m/s \\
$\sigma_u,\sigma_v,\sigma_w$ \> \hspace{2mm} = \hspace{2mm} turbulence standard deviations, m/s \\
$\mathrm{TI}$ \> \hspace{2mm} = \hspace{2mm} turbulence intensity, $\sigma_u/U_\infty$ \\
$L_u,L_v,L_w$ \> \hspace{2mm} = \hspace{2mm} integral length scales of turbulence, m \\
$\Phi_u,\Phi_w$ \> \hspace{2mm} = \hspace{2mm} one-dimensional turbulence power spectral densities, m$^2$/s \\
$w_m, d_m$ \> \hspace{2mm} = \hspace{2mm} 1-cosine discrete-gust peak velocity and length \\
$\Gamma_0, r_c$ \> \hspace{2mm} = \hspace{2mm} Oseen-vortex circulation, m$^2$/s; core radius, m \\
$\Delta U, \delta, U_f$ \> \hspace{2mm} = \hspace{2mm} gust-front velocity jump, m/s; thickness, m; propagation speed, m/s \\
$u_*, \kappa, z_0$ \> \hspace{2mm} = \hspace{2mm} friction velocity, m/s; von Kármán constant; roughness length, m \\
$\nu_t, C_s, \Delta$ \> \hspace{2mm} = \hspace{2mm} LES eddy viscosity, Smagorinsky constant, filter width \\
$\mathrm{Ma}$ \> \hspace{2mm} = \hspace{2mm} Mach number \\
$H$ \> \hspace{2mm} = \hspace{2mm} atmospheric scale height, m \\
\\
\emph{Vehicle and control quantities} \\
$\mathbf{p}, \mathbf{v}$ \> \hspace{2mm} = \hspace{2mm} vehicle position and velocity in the inertial frame \\
$\phi,\theta,\psi$ \> \hspace{2mm} = \hspace{2mm} roll, pitch, and yaw angles, rad \\
$\mathbf{u}$ \> \hspace{2mm} = \hspace{2mm} control input vector (collective thrust and moments) \\
$C_D, \omega_i$ \> \hspace{2mm} = \hspace{2mm} drag coefficient; rotor speeds, rad/s \\
$T_{\max}$ \> \hspace{2mm} = \hspace{2mm} maximum available collective thrust, N \\
$e$ \> \hspace{2mm} = \hspace{2mm} tracking error vector, $\mathbf{p}-\mathbf{p}_{\mathrm{ref}}$ \\
$R, \mathbf{\Omega}$ \> \hspace{2mm} = \hspace{2mm} body-to-inertial rotation matrix; body angular velocity, rad/s \\
$e_R, e_\Omega$ \> \hspace{2mm} = \hspace{2mm} geometric attitude and angular-velocity errors on SO(3) \\
$\gamma$ \> \hspace{2mm} = \hspace{2mm} PPO discount factor \\
\end{tabbing}

\section{Introduction}

Low-altitude aviation is undergoing rapid expansion~\cite{Huang25nature}. Atmospheric turbulence remains the dominant environmental disturbance for such operations, imposing unsteady loads that challenge both classical and learning-based control architectures.  Winds are neither stationary nor Gaussian in practice: coherent structures, shear-driven streaks, and intermittent gusts populate the atmospheric boundary layer, yet the vast majority of flight control simulations---and hence the training environments for learned policies---represent turbulence as a stationary, spectrally shaped random process, or worse, as independent white noise. The gap between the simplified wind models used for flight control training of drones and the physical complexity of real atmospheric flows is rarely treated as an independent variable, which motivates the current study.

For quadrotors, reinforcement-learning-based controllers have demonstrated agile and robust performance in both simulation and hardware deployments. At the larger end of the scale, online adaptive reinforcement learning flight control laws have been successfully flight-tested on CS-25-class transport aircraft~\cite{konatala2024flighttest}. Beyond standard architectures, hybrid evolutionary–reinforcement learning schemes have been developed for safety-informed fault-tolerant control, with training deliberately conducted on high-fidelity models to ensure reliability~\cite{gavra2024evolutionary}. For high-speed applications, reinforcement learning controllers for hypersonic vehicles have undergone structured performance comparisons against classical designs~\cite{wallace2024hypersonic}. At a methodological level, experience-replay mechanisms from deep reinforcement learning have been shown to strengthen the excitation conditions of neural-network adaptive control, thereby tightening the connection between learning-based and classical adaptive flight control~\cite{qu2025experiencereplay}.

As learned controllers move toward flight-test and certification contexts, the fidelity of the simulation environments in which they are trained becomes a prerequisite question rather than an implementation detail. {Evaluation methodology is evolving in response: deep reinforcement learning itself has been used to search out worst-case failure scenarios of complex nonlinear flight-control designs, complementing classical verification and validation~\cite{braun2023worstcase}, and learning-based adaptive-dynamic-programming control has reached wind-tunnel validation, alleviating gust-load peaks by morphing camber~\cite{zhang2026adpgust}.} Yet a recurring observation across the literature is that learned policies are exquisitely sensitive to the disturbances under which they are trained: a policy that is robust to the particular random process used during domain randomization may fail abruptly when the disturbance realization at test time differs in \emph{flow structure}, not merely in magnitude. Wind is the dominant external disturbance for low-altitude quadrotor operations, and it is almost universally modeled in the learning-control literature as a stationary, Gaussian, spectrally shaped random process---or, worse, as independent white noise on the force and torque residuals.

This sensitivity is not surprising when the environment under control {is} the atmosphere. In autonomous cross-country soaring, hierarchical reinforcement-learning policies succeed or fail according to how wind fields and thermal updrafts are represented to the agent~\cite{notter2023soaring}; for rotorcraft---the certified-aircraft category closest to the multirotor disturbance regime---gust reconstruction from vehicle response has been shown to hinge on closed-loop shaping of the vehicle--wind interaction~\cite{ozkurt2024gust}. In both lines of work the wind field itself is taken as given; its physical fidelity is not treated as an experimental variable.


Recent careful-comparison studies have established that conclusions about learned versus classical controllers depend strongly on simualtion and experimental protocols. Kunapuli et al.~\cite{kunapuli2025leveling} showed that, under matched conditions, classical and learned quadrotor trajectory-tracking controllers occupy a much narrower performance gap than commonly claimed, and that protocol details---task definition, tuning budget, disturbance injection---dominate the comparison. Chen et al.~\cite{chen2024whatmatters} systematically ablated the ingredients of a zero-shot sim-to-real reinforcement learning quadrotor policy, identifying observation design and dynamics randomization as decisive factors. Neither study treats the {physical fidelity of the wind field itself} as an independent variable: turbulence models of very different structure (discrete gusts, stationary stochastic spectra, coherent vortical structures, and turbulence-resolving simulation output) are usually conflated.

{Wind-disturbance modeling practice in the wider aerospace literature confirms the abovementioned gap from two directions. On the measurement side, the classical statistical gust models are now being calibrated directly against atmospheric data---the parameters of a three-dimensional gust model have been estimated from airborne wind-lidar measurements~\cite{musso2025gustlidar}, and the disturbances actually encountered by a large commercial aircraft have been reconstructed from flight data~\cite{hinson2026atmospheric}---and both studies quantify, case by case, how far realized atmospheric turbulence departs from the stationary Gaussian models used in simulation and training. On the estimation side, wind information is beginning to enter the control loop explicitly, adaptively and with uncertainty quantification in an empirical-Bayesian framework~\cite{kiehn2024empiricalbayes}, and with flight-test validation in model-based $H_\infty$ filtering~\cite{gahan2025hinftywind}; yet both lines of work still presuppose a modeled wind-field structure against which the estimate is computed.}

Wind-aware learning controllers have recently begun to emerge.  Tasim and Sun~\cite{tasim2026windaware} trained a wind-aware reinforcement learning policy with a learned onboard wind estimator in von K\'arm\'an turbulence; Wang et al.~\cite{wang2025piwan} proposed a physics-informed wind-adaptive network for quadrotor dynamics prediction under wind; Hsieh et al.~\cite{hsieh2025tightformations} addressed online adaptation for quadrotors flying in tight formations, where aerodynamic interaction dominates. Each of these works commits to a single disturbance model class, so the marginal value of flow fidelity---statistical turbulence versus coherent-structure-laden versus large eddy simulation (LES) derived fields---remains unquantified. The same gap exists in the guidance-and-control literature at large: recent work on learning-enhanced control~\cite{federici2026rlmpc,hirst2026learningmpc,martini2026koopman,liu2025nnincrem} demonstrates the maturity of learning-based methods across platforms, but evaluation disturbances are invariably simplified relative to the turbulence that the underlying aircraft actually traverse.


{A parallel development on the fluids side makes the fidelity question timely: flow fields of CFD quality are now routinely coupled to aerospace vehicle applications. Full-scale wind databases based upon reynolds-averaged Navier-Stokes equations over real urban geometry, compressed online by gappy proper orthogonal decomposition, already drive minimum-energy quadrotor trajectory planning~\cite{ebert2023windyurban}. At the gust-physics end, iterative maneuver optimization against high-fidelity simulation and experiment has clarified how a wing should respond to a transverse discrete-gust encounter~\cite{xu2023transversegust}, and wind-tunnel measurements of wings submerged in turbulent bluff-body wakes show that the ratio of coherent wake-eddy scale to wing span governs the unsteady loads~\cite{zhang2023bluffbodywake}---precisely the organized structure that stationary statistical turbulence does not contain by construction. Real-time data-driven adaptive lift regulation in developing long-duration gusts, demonstrated without any prior gust model~\cite{portelladelgado2025gusts}, completes the picture: controllers are now expected to cope with gusts whose structure is neither Gaussian nor stationary. What none of these works does is vary the fidelity of the disturbance field itself as an experimental variable; each commits to a single field generator, so the marginal value of physical fidelity remains unmeasured across the entire spectrum from discrete gusts to turbulence-resolving simulation output.}

To address the gap, this paper asks a deliberately narrow question with broad practical consequences: \emph{does the physical fidelity of the training wind field matter for the robustness and transferability of learned quadrotor controllers?} 
We should admit that the present study is deliberately simulator-based; its aim is to establish a protocol and to identify which fidelity level suffices, before flight-test resources are committed.  
{Precisely because the study is simulator-based, the rigor of the data used to develop and assess this question is treated as a first-class requirement: each disturbance generator is validated quantitatively against the analytical or certification-standard model it implements, the vehicle model is a peer-reviewed benchmark with documented system identification, cross-level data are matched on dimensionless grounds, and the assessment statistics are fixed a priori. }

The remaining structure of the present paper and the corresponding contributions are:
\begin{enumerate}
  \item A hierarchy of five wind-disturbance fidelity levels (L0--L4), from wind-free flight to LES-derived flow fields, with complete mathematical definitions and dimensionless matching criteria (turbulence intensity, integral-scale-to-airframe ratio, spectral shape) that allow fair cross-model comparison (Secs.~\ref{sec:wind}--\ref{sec:solver}).
  \item A full train-fidelity $\times$ test-fidelity cross-evaluation protocol applied to proximal policy optimization (PPO) controllers with a training-free proportional-integral-derivative (PID) reference under matched fairness conditions, with mechanism-level diagnostics (control-input spectra, saturation duty) in addition to performance metrics (Secs.~\ref{sec:control}--\ref{sec:protocol}).
\end{enumerate}
Next, Sec.~\ref{sec:discussion} presents the discussions of the results. 
Finally, Sec.~\ref{sec:conclusions} concluds the present study. 

\section{Wind Disturbance Models}\label{sec:wind}

We define five fidelity levels, all sharing a common interface: a callable field $\mathbf{w}(\mathbf{x},t) = (u,v,w)$. This field is evaluated at the vehicle's center of mass and, where applicable, at rotor disk points; the vehicle simulation subsequently converts these values into the relative wind used by its aerodynamic model.

The lower rungs of the hierarchy deliberately mirror the two-track treatment of atmospheric disturbances prescribed by airworthiness standards. Certification specifications such as European Union's CS-25.341 (gust and turbulence loads) pair deterministic discrete gusts---modeled as 1-cosine shapes---with continuous turbulence characterized by von K\'arm\'an spectra. This duality also underpins recent work on automated gust-load alleviation~\cite{cavaliere2024gust}. Our L1 and L2 levels are the direct counterparts of these two tracks, respectively.

The common interface itself is a modeling choice with consequences---the detail with which vehicle--wind coupling is represented has been shown to markedly affect model-based wind estimation from aircraft motion once unsteady aerodynamics is accounted for~\cite{halefom2024unsteady}, which is precisely why the conversion of $\mathbf{w}(\mathbf{x},t)$ into relative wind is held identical across fidelity levels rather than being allowed to co-vary with the field generator. Levels are matched in the dimensionless sense of Sec.~\ref{sec:protocol}: equal mean speed $U_\infty$, equal turbulence intensity $\mathrm{TI}$, and comparable integral scale ratio $L_u/d$, where $d$ is a characteristic airframe length.

\begin{figure}[!tbp]
  \centering
  \includegraphics[width=3.5in]{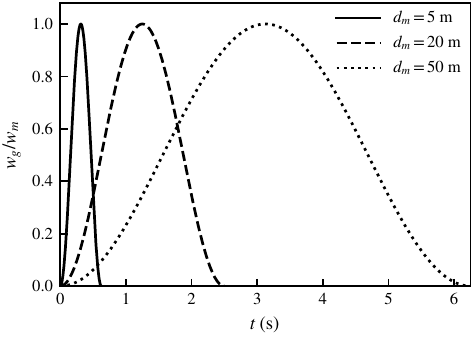}
  \caption{1-cosine gust time histories $w_g/w_m$ of Eq.~\eqref{eq:onecosine} at $U_\infty=8$\,m/s for gust lengths $d_m=5$, $20$, and $50$\,m, spanning the randomized range of the L1 sampler.}
  \label{fig:l1gust}
\end{figure}
\subsection{L0 --- Wind-free baseline}

{Figure~\ref{fig:l1gust} summarizes the key flow features of this level.}
$\mathbf{w}(\mathbf{x},t) \equiv 0$. This level isolates nominal tracking performance and provides the reference row and column of the cross-fidelity matrix.

\subsection{L1 --- Discrete deterministic gusts}

Following the flying-qualities specification MIL-F-8785C~\cite{milf8785c}, the discrete gust family is the 1-cosine shape
\begin{equation}
w_g(s) = \frac{w_m}{2}\left[1-\cos\!\left(\frac{2\pi s}{d_m}\right)\right], \qquad 0 \le s \le d_m,
\label{eq:onecosine}
\end{equation}
where $s$ is distance penetrated into the gust, $w_m$ the peak gust velocity, and $d_m$ the gust length. Pairs $(w_m, d_m)$ are selected to span the gust-gradient lengths to which the vehicle is most responsive (for a quadrotor, $d_m$ of order one to ten rotor diameters), and gusts are injected along each body axis independently and in combination. L1 is deterministic and reproducible, making it the cheapest probe of transient response.

\subsection{L2 --- Statistical turbulence}

\begin{figure}[!tbp]
  \centering
  \includegraphics[width=3.5in]{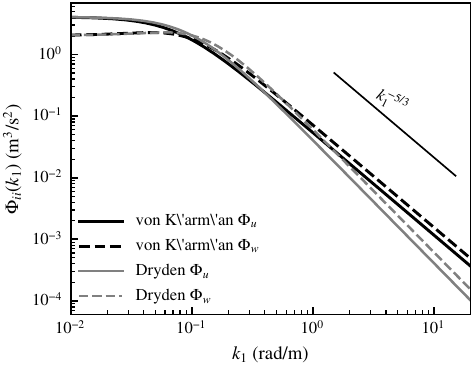}
  \caption{Von K\'arm\'an and Dryden one-dimensional spectra at $U_\infty=8$\,m/s, $\mathrm{TI}=10\,\%$, $L_u=10$\,m; the thin line marks the $-5/3$ inertial-range rolloff.}
  \label{fig:l2turb}
\end{figure}

{Figure~\ref{fig:l2turb} shows the spectral content that defines this level.} The stationary random component follows the Dryden and von K\'arm\'an models standardized for aircraft applications~\cite{dryden1943,vonkarman1948}. The Dryden model realizes turbulence by filtering white noise through rational transfer functions, e.g., for the axial component,
\begin{equation}
H_u(s) = \sigma_u\sqrt{\frac{2L_u}{\pi V}}\,\frac{1}{1+\left(L_u/V\right)s},
\end{equation}
with $V$ the vehicle airspeed. {The lateral and vertical components follow the corresponding second-order filters
\begin{equation}
H_{v,w}(s) = \sigma_{v,w}\sqrt{\frac{L_{v,w}}{\pi V}}\,
\frac{1+\sqrt{3}\,\left(L_{v,w}/V\right)s}{\left[1+\left(L_{v,w}/V\right)s\right]^2},
\label{eq:drydenvw}
\end{equation}
with the low-altitude prescriptions of MIL-F-8785C relating scales and intensities to height above the surface~\cite{milf8785c}.} The von K\'arm\'an model prescribes spectra with the physically correct $-5/3$ inertial-range rolloff,
\begin{equation}
\Phi_u(\Omega) = \sigma_u^2\,\frac{2L_u}{\pi}\,\frac{1}{\left[1+(1.339\,L_u\Omega)^2\right]^{5/6}},
\end{equation}
\begin{equation}
\Phi_w(\Omega) = \sigma_w^2\,\frac{L_w}{\pi}\,
\frac{1+\tfrac{8}{3}\,(1.339\,L_w\Omega)^2}{\left[1+(1.339\,L_w\Omega)^2\right]^{11/6}},
\end{equation}
where $\Omega$ is spatial frequency. Both models are stationary, Gaussian, and isotropic given scale ratios; they contain no coherent structure by construction. A mean shear is superposed via the power-law profile
\begin{equation}
U(z) = U_{\mathrm{ref}}\left(\frac{z}{z_{\mathrm{ref}}}\right)^{\alpha}, \qquad \alpha \approx \frac{1}{7},
\label{eq:powerlaw}
\end{equation}
representative of a neutral atmospheric surface layer. {The spectra are realized by Fourier mode superposition with amplitudes drawn from the above spectra. Throughout the sweep the integral scale is held fixed at $L_u=10$\,m---$L_u/d \approx 46$ for the racer airframe ($d=0.22$\,m tip-to-tip)---so that the scale ratio does not confound the fidelity variable, and intensities follow $\sigma_u=\mathrm{TI}\cdot U_\infty$.}

\subsection{L3 --- Synthetic coherent structures}

Real boundary-layer turbulence is intermittent and organized. L3 embeds parameterized coherent structures in an L2 background. The building block is the Oseen (Lamb--Oseen) vortex~\cite{lamb1932hydrodynamics} with tangential velocity
\begin{equation}
v_\vartheta(r,t) = \frac{\Gamma_0}{2\pi r}\left[1-\exp\!\left(-\frac{r^2}{r_c^2(t)}\right)\right],
\end{equation}
{advected past the vehicle with the mean wind; viscous core spreading, $r_c^2(t)=r_{c0}^2+4\nu t$, is negligible on episode timescales ($4\nu t/r_c^2 \sim 10^{-4}$) and $r_c$ is therefore held fixed. Gust fronts are modeled as translating tanh-smoothed velocity jumps,
\begin{equation}
\mathbf{w}_f(\mathbf{x},t) = \frac{\Delta U}{2}\left[1+\tanh\!\left(\frac{\mathbf{x}\cdot\hat{\mathbf{e}}_f-s_f(t)}{\delta}\right)\right]\hat{\mathbf{e}}_f, \qquad s_f(t)=s_0+U_f t,
\label{eq:gustfront}
\end{equation}
of amplitude $\Delta U$, thickness $\delta$, and propagation speed $U_f$ along the front normal $\hat{\mathbf{e}}_f$.} Circulations $\Gamma_0$ and core radii $r_c$ are sampled so that the induced velocity variance added to the L2 background preserves the target TI while shifting probability mass into rare, large, spatially correlated excursions---precisely the events that stationary Gaussian models under-represent.

\begin{figure}[!tbp]
  \centering
  \includegraphics[width=3.5in]{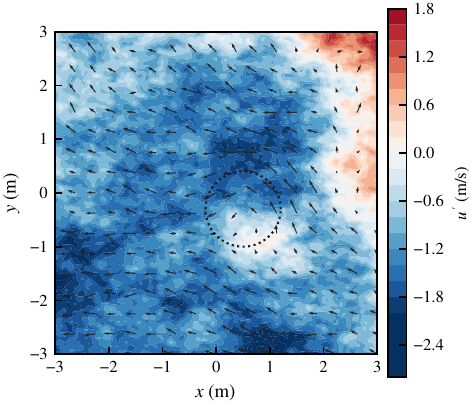}
  \caption{Oseen vortex embedded in an L2 statistical-background realization at the hover plane (streamwise velocity perturbation).}
  \label{fig:l3coherent}
\end{figure}

In aerodynamics the Oseen vortex model is widely applied to investigate aircraft wake-vortex evolution and its impact on following aircraft, turbulent mixing and energy cascade, and geophysical mesoscale eddies in oceans and atmospheres. Its analytical simplicity and physical fidelity make it a benchmark for understanding vortex dynamics across engineering and environmental fluid mechanics. 
{Figure~\ref{fig:l3coherent} makes the added structure visible.} The coherent, spatially correlated structure riding on the stochastic background is the defining content of this level, absent from L2 by construction.

\subsection{L4 --- LES-derived flow fields}

At the highest fidelity, three-dimensional, time-varying fields are obtained from large-eddy simulation of a boundary layer, in which the filtered incompressible Navier--Stokes equations
\begin{equation}
\frac{\partial \tilde{u}_i}{\partial x_i}=0, \qquad
\frac{\partial \tilde{u}_i}{\partial t} + \frac{\partial \tilde{u}_i\tilde{u}_j}{\partial x_j}
= -\frac{1}{\rho_0}\frac{\partial \tilde{p}}{\partial x_i}
+ \frac{\partial}{\partial x_j}\!\left[2(\nu+\nu_t)\tilde{S}_{ij}\right]
\end{equation}
are integrated with a Smagorinsky--Lilly subgrid closure $\nu_t = (C_s\Delta)^2\lvert\tilde{S}\rvert$~\cite{smagorinsky1963,deardorff1970}. {The simulation targets the neutral surface-layer log law
\begin{equation}
U(z) = \frac{u_*}{\kappa}\ln\!\left(\frac{z}{z_0}\right),
\label{eq:loglaw}
\end{equation}
with friction velocity $u_*$, von K\'arm\'an constant $\kappa=0.41$, and aerodynamic roughness length $z_0$~\cite{pope2000turbulent,moeng1984les}.} Snapshots are exported as space-time lookup tables and interpolated onto the vehicle trajectory. Unlike L2, the L4 field contains streaks, sweeps, ejections, and shear-consistent anisotropy; unlike L3, these structures emerge from dynamics rather than parameterization. The production record is generated once at $U_{\mathrm{ref}}=8$\,m/s; other mean-wind conditions are obtained by uniform velocity rescaling $\mathbf{w}'=\lambda\,\mathbf{w}$, $\lambda=U_\infty/U_{\mathrm{ref}}$, which---by the self-similarity of the neutral log law, in which every velocity scale is proportional to $u_*$---yields another neutral boundary-layer realization with unchanged $\mathrm{TI}$ and $L_u/d$. The same rescaling, sampled per episode over $\lambda\in[0,1]$, gives L4 training the same mean-wind coverage ($0$--$8$\,m/s) as the L1--L3 samplers. The solver and the validity of this choice are the subject of the next section.

\begin{figure}[!tbp]
  \centering
  \includegraphics[width=3.5in]{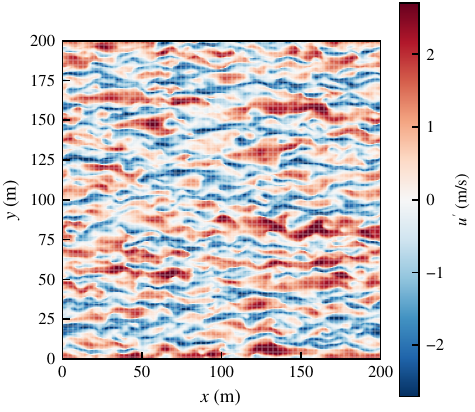}
  \caption{Instantaneous streamwise fluctuation $u'$ on the lowest exported slice ($z\approx1.2$\,m, $200\times200$\,m) of the production LES boundary layer.}
  \label{fig:l4les}
\end{figure}

{Figure~\ref{fig:l4les} illustrates the emergent structure of an L4 field. The flow exhibits a streaky, shear-consistent organization that arises naturally from the flow dynamics, in contrast to the parameterized coherent structures of Fig.~\ref{fig:l3coherent} and the structureless spectra of Fig.~\ref{fig:l2turb}.}

\section{High-Fidelity Flow Solver for Low-Altitude Air}\label{sec:solver}

In this work, the L4 fields are computed using Oceananigans.jl~\cite{ramadhan2020oceananigans}, a package originally developed for oceanic flow simulations. We adopt this solver primarily for its exceptional computational efficiency on graphical processor units (GPUs). This choice warrants explicit justification, which we provide on three grounds: the governing equations solved, quantitative scaling for low-altitude air, and precedent in the atmospheric-science literature. 

\subsection{The governing equations}

Oceananigans integrates the incompressible Navier--Stokes equations under the Boussinesq approximation: momentum is advected, diffused, and pressured-projected on a staggered finite-volume grid, while buoyancy enters only through a linear body-force term. The ocean-specific machinery---a seawater equation of state, active salinity, and Coriolis rotation appropriate to geophysical scales---consists of {optional, swappable components}. Setting the equation of state to constant density, deactivating tracers and rotation, and supplying an aerodynamic rough-wall boundary condition reduces the model to a canonical turbulent boundary-layer/channel LES solver for air. The governing equations are then adapted for low-altitude applications.

\subsection{The Boussinesq approximation at low altitude}

Three independent estimates bound the error of the Boussinesq/incompressible description for the present application:

\begin{enumerate}
  \item \emph{Depth ratio.} The flow of interest occupies a surface layer of depth $D \approx 50$--$100$\,m, whereas the atmospheric scale height is $H \approx 8.4$\,km. Hence $D/H \sim 10^{-2}$, and the hydrostatic density variation across the domain is of order one percent---negligible against turbulent velocity fluctuations of order $\mathrm{TI}\cdot U_\infty$.
  \item \emph{Mach number.} With wind speeds up to $U_\infty \approx 12$\,m/s and gust increments of similar magnitude, $\mathrm{Ma} \approx 0.04$, so compressibility effects scale as $\mathrm{Ma}^2 \approx 2\times10^{-3}$, which is negligible. 
  \item \emph{Buoyancy magnitude.} Density fluctuations entering the buoyancy term satisfy $\Delta\rho/\rho \approx \Delta T/T$; for temperature fluctuations of a few kelvin over $T\approx300$\,K, this is again at the percent level. Moreover, for the neutral stratification targeted here (strong-wind boundary layers are near-neutral by construction), the buoyancy term vanishes identically and the Boussinesq system degenerates exactly to the incompressible Navier--Stokes equations.
\end{enumerate}

In short, at the scales of quadrotor operations the distinction between ``ocean water'' and ``air'' enters only through parameter values (kinematic viscosity, roughness length, mean shear), not through the governing equations.

\subsection{Atmospheric large-eddy simulation}

The atmospheric community reached the same conclusion decades ago. The subgrid closures used here were invented for atmospheric boundary layers: Smagorinsky's nonlinear eddy viscosity in 1963~\cite{smagorinsky1963}, Deardorff's pioneering three-dimensional boundary-layer LES in 1970~\cite{deardorff1970}{, and Moeng's planetary-boundary-layer LES model that set the template for the modern atmospheric codes~\cite{moeng1984les}}. Contemporary atmospheric LES codes---PALM~\cite{maronga2020palm}, DALES~\cite{heus2010dales}, and the LES mode of WRF---are built on the Boussinesq or anelastic approximations; the review of Stoll et al.~\cite{stoll2020les} documents this consensus for boundary-layer flows. Most directly, the Climate Modeling Alliance team has announced Breeze.jl~\cite{breeze2026}, an atmosphere model built on the Oceananigans kernel targeting $O(10\,\mathrm{m})$ large-eddy simulations through mesoscale applications---the same solver, the same equations, atmospheric branding. Using Oceananigans itself for a neutral, low-altitude boundary layer is therefore a conservative rather than an exotic choice.

\subsection{Coupling architecture}

The LES is used strictly {offline} in this study. The flow solver produces a four-dimensional dataset $\mathbf{w}(\mathbf{x},t)$ once per configuration; the quadrotor simulation then samples this field by trilinear interpolation in space and linear interpolation in time, and adds the result to the vehicle velocity to form the relative wind seen by the rotor and body aerodynamic models. No two-way coupling is attempted: the quadrotor is a point/momentum-theory actor in the wind field, not an immersed boundary. This separation keeps the aerodynamic model inside the vehicle simulator, where it can be identical across fidelity levels, and makes L4 a drop-in replacement for L1--L3 behind the common $\mathbf{w}(\mathbf{x},t)$ interface.

Cross-level comparisons are meaningful only if the fields are matched on dimensionless grounds rather than on the identity of the generating fluid: turbulence intensity $\mathrm{TI}=\sigma_u/U_\infty$, integral-scale ratio $L_u/d$ relative to the airframe, and spectral shape (inertial-range exponent and anisotropy ratios) are the matching variables, with the parameter sweep of Sec.~\ref{sec:protocol} bracketing them.

\subsection{Configuration and validation}

The production L4 dataset is computed on a single consumer GPU (NVIDIA RTX 3090) at $256\times256\times128$ resolution ($\Delta \approx 0.78$\,m) over the $200\times200\times100$\,m domain, driven toward the neutral log law with $u_*=0.551$\,m/s, $z_0=0.03$\,m, and $U_{\mathrm{ref}}=8$\,m/s at $z_{\mathrm{ref}}=10$\,m. After a 250\,s spin-up (ten advection through-times $L_x/U_{\mathrm{ref}}$), 601 snapshots of the 1--20\,m activity band are exported at 0.5\,s intervals over a 300\,s sampling window. The wall-clock cost is approximately 1.2\,h for 550\,s of simulated time, about two orders of magnitude faster than the multithreaded-CPU estimate for the same cell-step count, which is what makes the train-fidelity $\times$ test-fidelity campaign of Sec.~\ref{sec:protocol} practical on desktop hardware. {A half-split convergence check on this original export (below) showed that a 300\,s window is too short for second-order statistics; the sampling window was therefore doubled. The identical configuration was re-integrated on 16 CPU threads ($\approx$\,20\,h wall clock) with a 400\,s spin-up (sixteen advection through-times), and 1201 snapshots were exported at 0.5\,s intervals over a 600\,s sampling window (simulation time $t=400$--$1000$\,s). All L4 training and evaluation results reported below use this extended 600\,s dataset.}

\begin{figure*}[!tbp]
  \centering
  \begin{subfigure}[t]{0.48\textwidth}
    \includegraphics[width=\textwidth]{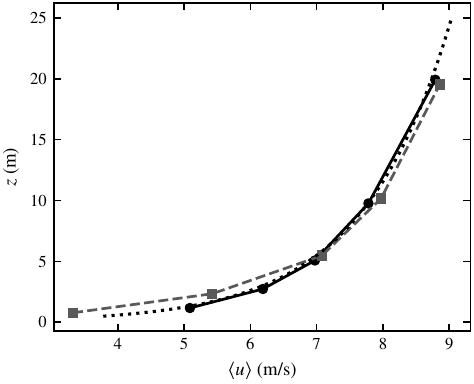}
    \caption{Solid with circles, 600\,s production run; dashed with squares, 300\,s reference run; dotted, target log law.}\label{fig:lesprofiles_b}
  \end{subfigure}\hfill
  \begin{subfigure}[t]{0.48\textwidth}
    \includegraphics[width=\textwidth]{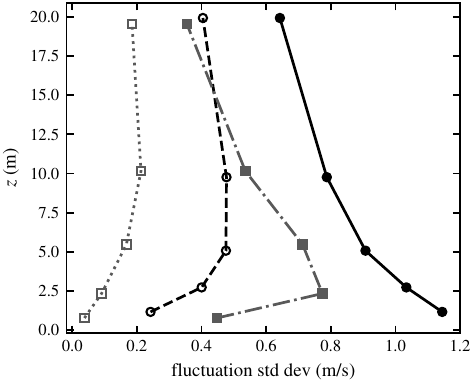}
    \caption{Black, production grid; gray, reference grid; solid/filled, $\sigma_u$; dashed/open, $\sigma_w$.}\label{fig:lesprofiles_c}
  \end{subfigure}
  \caption{LES validation against the target neutral log law: (a) time-averaged streamwise velocity and (b) fluctuation standard-deviation profiles.}
  \label{fig:lesprofiles}
\end{figure*}
{Momentum advection uses a fifth-order WENO scheme on the staggered grid, and the Smagorinsky--Lilly closure uses $C_s=0.16$ with filter width $\Delta=(\Delta x\,\Delta y\,\Delta z)^{1/3}$. The boundary layer is driven by relaxing the horizontal-mean streamwise velocity toward the log-law profile of Eq.~\eqref{eq:loglaw} with a time constant of 30\,s, and the bottom boundary imposes quadratic drag consistent with the same log law evaluated at the first grid level; periodic conditions apply in both horizontal directions. The exported dataset comprises the three velocity components on five horizontal slices spanning the 1--20\,m flight band ($z\approx1.2,2.7,5.1,9.8,19.9$\,m), stored in single precision; between slices the vehicle simulation interpolates linearly in the vertical. At evaluation time the record is replayed on a loop with a per-episode randomized phase, so that successive episodes sample uncorrelated segments of the field.}

Table~\ref{tab:lesvalid} and Fig.~\ref{fig:lesprofiles} validate the exported field against the imposed log law. The {600\,s} time-averaged mean profile reproduces the target within {$3.3\,\%$} at every sampled height; the turbulence intensity at the 10\,m reference height is {$\mathrm{TI}=10.1\,\%$, essentially coinciding with} the $\mathrm{TI}=10\,\%$ nominal evaluation condition of Sec.~\ref{sec:protocol}; and the fluctuation field is strongly anisotropic ($\sigma_w/\sigma_u \approx 0.6$ at $z\approx10$\,m), a shear-consistent property that the isotropic L2 construction lacks by design. A $128\times128\times64$ GPU reference run at identical $u_*$ and $z_0$ confirms that the {fine and coarse grids} converge to the same mean state, while the finer {production} grid resolves substantially more near-wall fluctuation energy (Fig.~\ref{fig:lesprofiles}b)---the coherent-structure content that motivates L4 in the first place.

{Statistical sufficiency of the record was quantified by a half-split test. Differencing the 400--700\,s and 700--1000\,s halves, the horizontal-mean profiles agree to within $2.1\,\%$ at all heights, and at the 1\,m hover altitude the second-order moments agree to $\approx5\,\%$ ($\sigma_u$: $5.0\,\%$, $\sigma_w$: $4.4\,\%$, TI: $3.0\,\%$). The integral time scale of $u$ at $z\approx1.2$\,m is $T_{\mathrm{int}}\approx3.6$\,s, so the 600\,s window contains $\approx165$ independent samples of the local turbulence. The residual half-to-half spread in $\sigma_u$ grows with height---up to $\approx29\,\%$ at $z\approx20$\,m---reflecting the slowly relaxing outer-layer modes of the 30\,s profile relaxation; these levels lie well above the flight band and do not enter the hover-task dynamics. Doubling the window from 300\,s to 600\,s reduced the worst flight-band discrepancy from $\approx40\,\%$ to $\approx5\,\%$, which is why the extended dataset was adopted for all reported L4 results.}

\begin{table}[!htb]
  \caption{Validation of the production LES against the imposed neutral log law ($u_*=0.551$\,m/s, $z_0=0.03$\,m): time-averaged mean velocity, relative deviation, and fluctuation standard deviations.}
  \label{tab:lesvalid}
  \centering
  \small
  \begin{tabular}{@{}ccccc@{}}
    \toprule
    $z$ [m] & $\langle u\rangle$ [m/s] & log law [m/s] & dev. [\%] & $\sigma_u$ / $\sigma_w$ [m/s]\\
    \midrule
    1.17  &  {5.09} & {4.92} & {$+3.3$} & {1.15 / 0.24}\\
    2.73  & {6.19} & {6.06} & {$+2.1$} & {1.03 / 0.40}\\
    5.08  & {6.97} & {6.89} & {$+1.2$} & {0.91 / 0.48}\\
    9.77  & {7.78} & {7.77} & {$+0.1$} & {0.79 / 0.48}\\
    19.92 & {8.79} & {8.73} & {$+0.7$} & {0.64 / 0.40}\\
    \bottomrule
  \end{tabular}
\end{table}

\section{Control Architectures}\label{sec:control}

{Three} controller families are evaluated: {two training-free classical baselines---a cascaded PID controller and a geometric SE(3) tracking controller---}and learned PPO reinforcement learning policies, spanning the fixed-gain versus offline-trained ends of the design spectrum.

\begin{figure}[!tbp]
  \centering
  \includegraphics[width=3.2in]{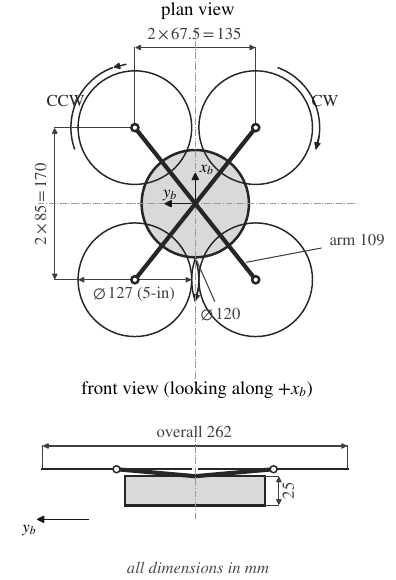}
  \caption{Racing-class quadrotor platform: X-configuration, 5-inch propellers, diagonally opposite rotors co-rotating; dimensions in mm, $m=0.83$\,kg.}
  \label{fig:platform}
\end{figure}

The simulated vehicle is the racing-class quadrotor model shipped with gym-pybullet-drones (mass 0.83\,kg, arm length 0.109\,m, thrust-to-weight ratio 4.17, maximum speed 55.6\,m/s; Fig.~\ref{fig:platform}), flying a stationary-hover task at 1\,m altitude with a 30\,Hz control loop; wind enters the dynamics through the relative-velocity drag model described in Sec.~\ref{sec:wind}. A companion campaign on the 27\,g Crazyflie-class cf2x model (thrust-to-weight ratio 2.25, maximum speed 8.3\,m/s) under the identical protocol provides the platform-sensitivity contrast reported in Sec.~\ref{sec:results}.

{The racing-class quadrotor descends from the first-person-view (FPV) drone-racing scene that crystallized in the mid-2010s around 5-inch-propeller, carbon-fiber X-frame machines built for agility and crash tolerance. The defining design choice of the class is excess control authority: four 5-inch propellers on roughly 0.2\,m motor spacing deliver thrust-to-weight ratios above four, far beyond photography-oriented platforms. A stiff, minimalist frame with centrally concentrated mass keeps the inertia tensor small and nearly isotropic, so the attitude bandwidth sits well above the dominant timescales of atmospheric gusts. The \texttt{racer.urdf} model shipped with gym-pybullet-drones~\cite{panerati2021learning} parameterizes exactly this class (Fig.~\ref{fig:platform}): 0.83\,kg mass, 0.109\,m arms, thrust-to-weight ratio 4.17, and a 55.6\,m/s speed envelope. It therefore represents the high-authority extreme of the platform-sensitivity axis studied here, complementing the 27\,g cf2x at the wind-sensitive end.}

\subsection{Cascaded PID baseline}

In drone applications, 
{the PID controller remains the reference architecture of practical flight control: three decades of industrial practice and theory~\cite{astrom1995pid} have made the cascaded PID loop the default against which learned controllers are judged{, and PID-based designs with full stability analysis remain an active subject of the guidance-and-control literature~\cite{lu2026wavepid}}.} The baseline is the differential-flatness cascaded PID controller (DSLPIDControl) shipped with the gym-pybullet-drones environment~\cite{panerati2021learning}: an outer position loop generating desired accelerations, mapped through flatness to desired attitudes and thrusts, with an inner attitude-rate loop. In the following comparison studies, gains are tuned once on the L0 task and held fixed for all wind conditions deliberately, since gain-scheduling on wind would itself constitute wind awareness.

\subsection{{Geometric control baseline}}

{Geometric control develops control laws directly on the configuration manifold of the rigid body rather than on local coordinate charts. The attitude of a quadrotor evolves on the special orthogonal group $SO(3)=\{R\in\mathbb{R}^{3\times3}: R^\top R=I,\ \det R=1\}$, the set of rotation matrices; appending the position gives the full pose on the special Euclidean group $SE(3)=SO(3)\ltimes\mathbb{R}^{3}$, the semidirect product of rotations and translations. Working intrinsically on these manifolds avoids the singularities of Euler-angle charts and the unwinding ambiguity of quaternion-coordinatized designs, and it yields stability guarantees that hold over almost the entire state space rather than in a local basin. The formulation remains an active front in the guidance-and-control literature: recent work includes adaptive tracking control formulated on the tangent bundle of $SE(3)$~\cite{seo2023se3}, semiglobally exponentially stable pose tracking in the dual-quaternion parameterization of $SE(3)$~\cite{zinage2024dualquat}, and H\"older-continuous finite-time-stable attitude tracking on $SO(3)$~\cite{wang2026holder}. The baseline adopted here is the classical member of this family.}

{The second training-free baseline is the geometric tracking controller of Lee, Leok, and McClamroch~\cite{lee2010geometric}, the reference nonlinear design for quadrotors, in which attitude errors are computed directly on $SO(3)$ rather than from Euler-angle decompositions, endowing the zero-hover equilibrium with almost-global exponential stability. With position and velocity errors $\mathbf{e}_x=\mathbf{p}-\mathbf{p}_{\mathrm{ref}}$ and $\mathbf{e}_v=\mathbf{v}-\mathbf{v}_{\mathrm{ref}}$, the desired force vector
\begin{equation}
\mathbf{F}_d = -k_x \mathbf{e}_x - k_v \mathbf{e}_v + m g\,\hat{\mathbf{e}}_3
\label{eq:geomforce}
\end{equation}
fixes the collective thrust $f=\mathbf{F}_d\cdot(R\hat{\mathbf{e}}_3)$ and the desired body axis $\mathbf{b}_{3,d}=\mathbf{F}_d/\lVert\mathbf{F}_d\rVert$; completing $R_d\in SO(3)$ with the reference yaw $\psi_{\mathrm{ref}}=0$, the attitude and angular-velocity errors
\begin{equation}
e_R = \tfrac{1}{2}\left(R_d^\top R - R^\top R_d\right)^{\vee}, \qquad
e_\Omega = \mathbf{\Omega} - R^\top R_d \mathbf{\Omega}_d
\label{eq:geomerr}
\end{equation}
define the control moment $\mathbf{M} = -k_R e_R - k_\Omega e_\Omega$, and $(f,\mathbf{M})$ are mapped to the four rotor speeds through the quadrotor mixer. The gains $(k_x,k_v,k_R,k_\Omega)$ per platform are tuned once on the L0 hover task and then frozen for all wind conditions, subjecting the geometric controller to the same fixed-gain discipline as the PID baseline.}

\subsection{PPO reinforcement learning}

The learned controllers are trained with PPO~\cite{schulman2017ppo} as implemented in Stable-Baselines3~\cite{raffin2021sb3}, on gym-pybullet-drones hover and trajectory-tracking tasks. {PPO maximizes a clipped surrogate objective that penalizes policy ratios outside $[1-\epsilon,1+\epsilon]$, retaining trust-region-like update stability with first-order cost~\cite{schulman2017ppo}; it is the de facto standard on-policy algorithm in the quadrotor reinforcement learning literature this paper draws on.} One policy is trained per fidelity level L0--L4 (five levels $\times$ five seeds), with observation, action, and reward definitions held constant across levels---only the disturbance field changes. Policies and value functions are two-layer $[256,256]$ multilayer perceptrons acting at the four-motor RPM level (full control authority), trained for $10^6$ environment steps per run with parallel rollout workers; wind conditions are re-sampled at every episode reset (domain randomization over the parameter grid of Sec.~\ref{sec:protocol}), so no policy can memorize a single wind realization.

{Three considerations motivate PPO over alternative reinforcement-learning algorithms for this study. First, its on-policy clipped-surrogate update tolerates the non-stationarity induced by per-episode wind re-sampling: off-policy actor--critic methods (SAC, TD3) can suffer value-extrapolation errors when a single replay buffer mixes heterogeneous disturbance regimes, whereas PPO's trust-region-like constraint keeps each policy update close to the behavior policy that generated the data~\cite{schulman2017ppo}. Second, its first-order cost and compatibility with massively parallel rollouts keep the present campaign---25 training runs (five levels $\times$ five seeds) plus 7290 evaluation episodes---tractable on desktop hardware; TRPO offers comparable stability at second-order cost per update. Third, recent guidance-and-control practice has converged on policy-gradient methods of this family as the default for continuous flight control, from flight-tested reinforcement learning adaptive laws on transport-class aircraft~\cite{konatala2024flighttest,gavra2024evolutionary} to meta-reinforcement-learning rocket landing guidance~\cite{carradori2026metarl} and model-based reinforcement learning control of strongly disturbed unsteady flows~\cite{liu2025mbrl}; PPO is therefore the representative, literature-anchored choice for the learned end of the design spectrum.}

\subsection{Fairness principles}

The comparison protocol adopts the fairness principles articulated by Kunapuli et al.~\cite{kunapuli2025leveling}: identical tasks, initial-condition distributions, and evaluation episodes for all controllers; a documented and comparable tuning/optimization budget for each family ({the PID and geometric gain searches are both single-platform, single-task procedures reported above, and }the PPO hyperparameter budget {is} reported); no controller receives information unavailable to the others (in particular, the reinforcement learning policies are not given privileged wind observations); and failure modes are reported alongside average performance. Benchmarking learned controllers against a classical baseline under common, documented conditions follows the performance-evaluation paradigm for reinforcement-learning flight control established {in the  literature}~\cite{wallace2024hypersonic}. The goal is not to rank controller families but to measure how {each} family responds to training-fidelity changes.


{Wind enters the quadrotor dynamics as a relative-velocity disturbance (Sec.~\ref{sec:protocol}), and each controller family has a characteristic way of accounting for it when the fixed-gain discipline of this study is relaxed:}
{\begin{enumerate}
  \item \emph{Cascaded PID.} Sustained mean wind produces a steady drag force that a pure position loop rejects only through integral action; gust transients additionally demand attitude bandwidth that calm-air gain tuning does not provide. The classical remedies are integral compensation of the steady offset, feedforward of a measured or estimated wind velocity, and gain scheduling on wind condition---each of which presupposes wind information that the fixed-gain baseline of this study deliberately denies itself (Sec.~\ref{sec:control}). Recent PID-based designs with full stability analysis~\cite{lu2026wavepid} illustrate how much structure must be added before a PID loop ceases to be the fragile object seen in the results below.
  \item \emph{Geometric SE(3) control.} The almost-global stability of the $SO(3)$ error dynamics of Eq.~\eqref{eq:geomerr} survives wind automatically---bounded disturbances degrade the tracking error gracefully rather than destroying it---but steady drag again leaves a bias. The geometric remedy augments the desired force of Eq.~\eqref{eq:geomforce} with a disturbance estimate, $-\hat{\mathbf{d}}$ for an estimated force disturbance $\hat{\mathbf{d}}$, or adds integral action on the manifold; adaptive variants estimate the disturbance online without leaving the $SE(3)$ framework~\cite{seo2023se3}. The baseline here omits such augmentation so that both classical controllers face the wind with the same fixed-gain discipline.
  \item \emph{Learned PPO policies.} A learned controller cannot be handed a wind model at deployment; wind awareness must be built in at training time. The available levers are (i) domain randomization over wind conditions and wind models---the lever studied in this paper---(ii) observation augmentation with an onboard wind estimate~\cite{tasim2026windaware,kiehn2024empiricalbayes}, and (iii) reward shaping that prices control effort against gust rejection. The fidelity hierarchy L0--L4 parametrizes lever (i); the results below quantify how much the structural realism of the randomization actually buys.
\end{enumerate}}

\section{{Evaluation and Results}}\label{sec:results}
\label{sec:protocol}

\subsection{{Data  rigor}}\label{sec:datarigor}

{The research question of this paper is empirical, and the data used to develop (train) and assess (test) it are synthetic by construction; their rigor is therefore established explicitly, as a chain of four links, before any controller result is presented.}

{\emph{(i) Disturbance generators against their references.} Every synthetic wind field is verified against the analytical or certification-standard model it implements, with quantitative tolerances, before entering the training and evaluation pools: the L1 gust histories reproduce the MIL-F-8785C 1-cosine definition of Eq.~\eqref{eq:onecosine} exactly; the L2 realizations are checked against the theoretical von K\'arm\'an and Dryden one-dimensional spectra of Sec.~\ref{sec:wind} under the Taylor hypothesis, over eight independent $600$\,s realizations, with integrated variance matching $\sigma_u^2=(\mathrm{TI}\,U_\infty)^2$ to within $0.3\,\%$, the inertial-range logarithmic slopes recovered ($-1.64$ to $-1.66$ against $-5/3$, and $-1.94$ to $-1.96$ against $-2$, respectively), and the superposed power-law shear profile recovered to within $0.6\,\%$ at the probed heights; and the L3 vortex component reproduces the Lamb--Oseen tangential profile to numerical precision, including the core value $v_\vartheta(r_c)=\Gamma_0(1-e^{-1})/(2\pi r_c)$. The L4 data carry the solver-level validation of Sec.~\ref{sec:solver}: agreement with the imposed neutral log law within $3.3\,\%$ at every exported height (Table~\ref{tab:lesvalid}), convergence against an independent coarse-grid run, and a half-split test establishing statistical sufficiency of the $600$\,s record ($\approx165$ independent integral-scale samples in the flight band); end-to-end replay of the exported slices---spatial and temporal interpolation, temporal periodicity, and per-seed reproducibility---is covered by automated pipeline assertions.}

{\emph{(ii) Vehicle model.} The quadrotor dynamics, actuator limits, and drag coefficients are those of the gym-pybullet-drones environment~\cite{panerati2021learning}, a peer-reviewed and openly maintained benchmark whose system identification against hardware is documented in the source publication. The present work modifies exactly one quantity---the air velocity entering the drag model of Eq.~\eqref{eq:drag}---confined to a single environment subclass, so the development and assessment data share one and the same dynamics code path.}

{\emph{(iii) Comparability across fidelity levels.} Cross-level data are matched on dimensionless grounds---equal mean speed $U_\infty$, equal turbulence intensity $\mathrm{TI}$, and comparable integral-scale ratio $L_u/d$ (Secs.~\ref{sec:wind} and~\ref{sec:solver})---so that the structural fidelity of the disturbance, and not a confounded intensity or scale, is the quantity varied between matrix cells.}

{\emph{(iv) Assessment statistics.} The evaluation campaign comprises 7290 episodes over 27 test conditions, aggregated as five seeds $\times$ ten episodes per cell; termination criteria are fixed a priori and identical for all controllers, training convergence is verified on held-out evaluation curves, and claims are graded against the seed-level spread as defined by the statistics protocol below. All generators, samplers, and analysis scripts form a fixed-seed pipeline to be released upon publication, so that every reported number is exactly reproducible.}

\subsection{{Simulation and wind-field integration}}

{All vehicle simulations run in gym-pybullet-drones~\cite{panerati2021learning}: wind physics lives entirely in a subclass. The environment \texttt{WindHoverAviary} (file \texttt{wind\_aviary.py}) subclasses the stock \texttt{HoverAviary} and injects an arbitrary wind field by evaluating the aerodynamic drag on the {relative} air velocity---the gust-corrected version of the simulator's drag model---applied at the center of mass at every PyBullet physics step,
\begin{equation}
\mathbf{F}_{\mathrm{drag}} = -C_{D}\,\Bigl({\textstyle\sum_{i=1}^{4}}\omega_i\Bigr)\,R^\top\bigl(\mathbf{v}-\mathbf{w}(\mathbf{x},t)\bigr),
\label{eq:drag}
\end{equation}
with $C_D$ the identified drag coefficient, $\omega_i$ the rotor speeds, $R$ the body attitude, and $\mathbf{w}(\mathbf{x},t)$ the wind field sampled at the vehicle position. The gymnasium interface (observation/action spaces, step/reset signatures) is unchanged, so the environment feeds directly into Stable-Baselines3 PPO. The field itself implements a common \texttt{WindField.velocity(x,\,t)} interface (\texttt{wind\_field.py}): the L1 1-cosine gusts, L2 von K\'arm\'an Fourier-superposition turbulence, and L3 Oseen-vortex/gust-front composites are analytic generators evaluated on the fly, while L4 (\texttt{LESWindField}, \texttt{les\_wind\_field.py}) replays the exported LES slices of Sec.~\ref{sec:solver}---trilinear in the horizontal, linear across the five exported $z$-levels, linear in time, looped with a per-episode randomized phase. A \texttt{WindSampler} (\texttt{wind\_samplers.py}) re-draws the wind condition at every episode reset (domain randomization over the grid of the parameter sweep below), so training and evaluation share exactly one code path: the fidelity level selects the generator, nothing else. The control loop runs at 30\,Hz on top of the finer PyBullet physics integration, and the identical relative-wind chain serves the PID, geometric, and PPO controllers.}


Every reinforcement learning policy is trained under exactly one fidelity level and evaluated under all five, producing a $5\times5$ train-fidelity $\times$ test-fidelity matrix per task and per metric. The diagonal measures in-distribution performance; off-diagonal entries measure transfer. The classical PID {and geometric control} baselines, which require no training, occupy {one row each, }evaluated across all five test levels and {serving} as fidelity-independent references.


Within each fidelity level, wind conditions are swept over mean speed $U_\infty \in \{0, 4, 8, 12\}$\,m/s and turbulence intensity $\mathrm{TI} \in \{5, 10, 20\}\,\%$, with integral scales matched across levels per Sec.~\ref{sec:solver}. L1 gust magnitudes are calibrated to the same $(U_\infty, \mathrm{TI})$ grid through the peak-to-mean ratio of the 1-cosine family.


Performance metrics are position and attitude RMSE against the reference trajectory, peak deviation, and crash rate (termination by ground contact, attitude excursion beyond $60^\circ$, or loss of containment). Mechanism-level diagnostics are (i) the power spectral density of the control inputs, exposing the frequency bands in which each policy fights the wind; (ii) thrust-saturation duty, the fraction of time at the thrust limit $T_{\max}$, quantifying margin consumption; and (iii) attitude-limit excursion frequency. Sample efficiency (environment steps to threshold performance) is recorded during training.


All experiments use five independent random seeds, and every matrix cell aggregates five seeds $\times$ ten episodes ($n=50$). {The analysis discipline is stated explicitly, because it bounds what the data can support. Termination criteria (ground contact, attitude excursion beyond $60^\circ$, loss of containment) and episode length are fixed a priori and identical for every controller, so no cell can improve its statistics by truncation. Matrix entries report mean and standard deviation across seeds and episodes. With a five-seed population, formal null-hypothesis tests would be underpowered, and claims are therefore graded by effect size: crash-rate differences are read against their binomial standard error ($\sqrt{p(1-p)/n}\approx4\,\%$ at $p=0.1$, $n=50$), and pairwise claims of the form ``training under $L_i$ transfers better to $L_j$ than training under $L_k$'' are made only where the difference of seed means exceeds the seed-level standard deviation; differences within that spread are reported as flat---which is itself among the findings below.}

Results are reported along three axes: the cross-fidelity performance matrix (Fig.~\ref{fig:matrix}, Table~\ref{tab:summary}), robustness boundaries in mean wind speed (Fig.~\ref{fig:boundary}), and frequency-domain mechanism analysis (Fig.~\ref{fig:spectrum}), followed by a platform-sensitivity contrast against the companion cf2x campaign{; Figs.~\ref{fig:compare}--\ref{fig:controlwind} give trajectory-level illustrations of the three controller families against the flow fields they fly through}. All entries aggregate five seeds $\times$ ten episodes. The total evaluation campaign comprises {7290} episodes over 27 test conditions.

\subsection{The cross-fidelity evaluation matrix}

\begin{figure*}[!tbp]
  \centering
  \begin{subfigure}[t]{0.48\textwidth}
    \centering
    \includegraphics[width=\textwidth]{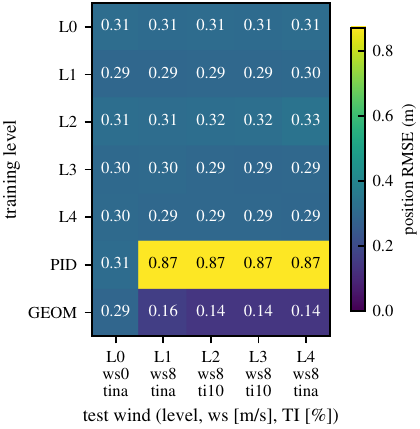}
    \caption{}\label{fig:matrix_rmse}
  \end{subfigure}\hfill
  \begin{subfigure}[t]{0.48\textwidth}
    \centering
    \includegraphics[width=\textwidth]{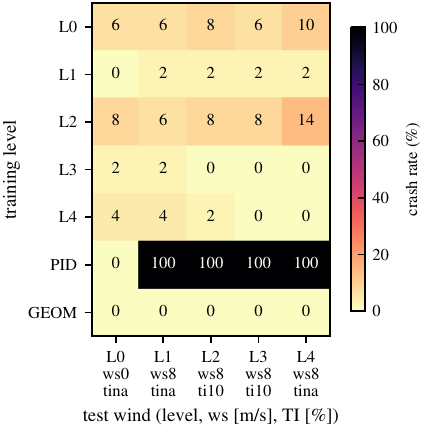}
    \caption{}\label{fig:matrix_crash}
  \end{subfigure}
  \caption{Cross-fidelity matrix: (a) position-tracking RMSE and (b) crash rate for PPO policies trained at level $L_i$ and tested at level $L_j$, each column at its representative test condition.} 
  \label{fig:matrix}
\end{figure*}
\begin{table}[!htb]
  \caption{Aggregate cross-fidelity summary at each level's representative condition: position RMSE (mean $\pm$ std), crash rate in parentheses (five seeds $\times$ ten episodes per cell).}
  \label{tab:summary}
  \centering
  \small
\begin{tabular}{lccccc}
\toprule
Training level & test L0 & test L1 & test L2 & test L3 & test L4 \\
\midrule
PPO (L0) & $0.31 \pm 0.06$ (6\%) & $0.31 \pm 0.04$ (6\%) & $0.31 \pm 0.05$ (8\%) & $0.31 \pm 0.04$ (6\%) & $0.31 \pm 0.05$ (10\%) \\
PPO (L1) & $0.29 \pm 0.02$ (0\%) & $0.29 \pm 0.03$ (2\%) & $0.29 \pm 0.03$ (2\%) & $0.29 \pm 0.03$ (2\%) & $0.30 \pm 0.03$ (2\%) \\
PPO (L2) & $0.31 \pm 0.06$ (8\%) & $0.31 \pm 0.05$ (6\%) & $0.32 \pm 0.06$ (8\%) & $0.32 \pm 0.06$ (8\%) & $0.33 \pm 0.08$ (14\%) \\
PPO (L3) & $0.30 \pm 0.04$ (2\%) & $0.30 \pm 0.07$ (2\%) & $0.29 \pm 0.01$ (0\%) & $0.29 \pm 0.01$ (0\%) & $0.29 \pm 0.01$ (0\%) \\
PPO (L4) & $0.30 \pm 0.06$ (4\%) & $0.29 \pm 0.02$ (4\%) & $0.29 \pm 0.02$ (2\%) & $0.29 \pm 0.01$ (0\%) & $0.29 \pm 0.01$ (0\%) \\
PID (baseline) & $0.31 \pm 0.00$ (0\%) & $0.87 \pm 0.01$ (100\%) & $0.87 \pm 0.01$ (100\%) & $0.87 \pm 0.01$ (100\%) & $0.87 \pm 0.01$ (100\%) \\
Geometric SE(3) (baseline) & $0.29 \pm 0.00$ (0\%) & $0.16 \pm 0.00$ (0\%) & $0.14 \pm 0.00$ (0\%) & $0.14 \pm 0.00$ (0\%) & $0.14 \pm 0.00$ (0\%) \\
\bottomrule
\end{tabular}

\end{table}
Three observations stand out in Fig.~\ref{fig:matrix} and Table~\ref{tab:summary}. First, the matrix is remarkably flat: across all twenty train--test pairs, position RMSE varies only between 0.29 and {0.33}\,m and crash rate between 0 and {14}\,\%. No diagonal (same-fidelity) advantage is visible, and even the L0-trained policy, which has never experienced wind during training, transfers to LES turbulence at 8\,m/s with a {10}\,\% crash rate. Second, the ranking that does exist is consistent and counterintuitive: the L1-trained policy (discrete-gust domain randomization) attains the lowest crash rate in every test column (0--2\,\%), closely followed by the L3- and L4-trained policies ({0--2}\,\% under the L2--L4 columns). Training on realistic turbulence is thus not necessary for robustness to it; cheap gust randomization is at least as good on this platform. Third, the training-free PID row delimits the value of learning: in calm air PID is as accurate as any learned policy (RMSE 0.31\,m, no crashes), but it fails catastrophically (100\,\% crash rate) in every windy test column. As discussed in Sec.~\ref{sec:discussion}, this failure is dominated by the transfer of the shipped gain schedule to the racer platform and should be read as a fixed-gain reference, not a tuned-best baseline. 

{The geometric SE(3) row sharpens exactly this point: with gains tuned once on the racer's L0 hover task, the geometric controller survives {every} test column at 0\,\% crash rate and posts the lowest RMSE of any row in every windy column (0.14--0.16\,m, against 0.29--0.33\,m for the learned policies). Two fixed-gain classical controllers therefore bracket the outcome: transferred gains fail outright, platform-tuned gains outperform the learned policies on accuracy. One metric asymmetry must be flagged: geometric episodes terminate upon reaching a $10^{-4}$\,m neighborhood of the target, so in calm air the climb transient dominates its RMSE (0.29\,m), while in wind the episodes run full length and the reported RMSE reflects its centimetre-level steady-state hover; the windy-column comparison is between full-length episodes for all rows.}

\subsection{Robustness of the controllers}

\begin{figure}[!tbp]
  \centering
  \includegraphics[width=5in]{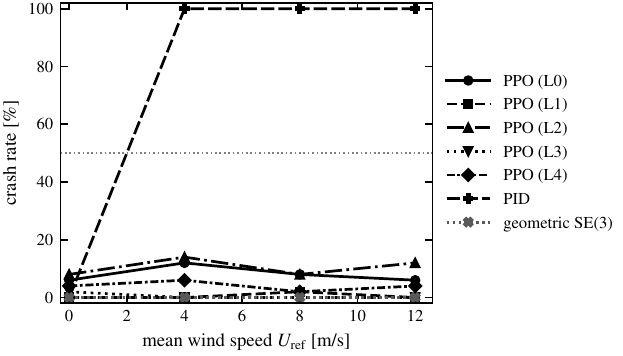}
  \caption{Crash rate versus mean wind speed under L2 statistical turbulence at $\mathrm{TI}=10\,\%$ for policies trained at each fidelity level and the classical baselines.}
  \label{fig:boundary}
\end{figure}
Figure~\ref{fig:boundary} shows crash rate versus mean wind speed under L2 turbulence at $\mathrm{TI}=10\,\%$. No PPO policy crosses the 20\,\% crash threshold anywhere in the swept envelope; the worst excursion is 14\,\% (L2-trained at 4\,m/s), and at 12\,m/s---a wind speed beyond the physical speed limit of the cf2x platform---all learned policies remain at or below 12\,\% under every test level, including L4 LES turbulence (0--8\,\%). The PID baseline crosses 50\,\% already at 2\,m/s{, while the geometric SE(3) baseline never crosses it---its crash count is zero at every wind speed}. Robustness on this platform is therefore not limited by the aerodynamic envelope within the certification-relevant wind range; the binding constraint visible in the cf2x campaign has disappeared.


\subsection{Control effort is spectrally insensitive to training fidelity}

\begin{figure*}[!tbp]
  \centering
  \begin{subfigure}[t]{0.48\textwidth}
    \centering
    \includegraphics[width=\textwidth]{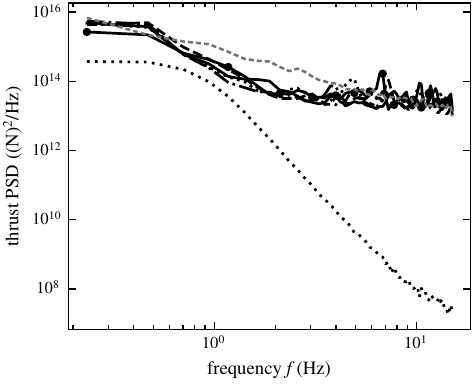}
    \caption{}\label{fig:psd_l2}
  \end{subfigure}\hfill
  \begin{subfigure}[t]{0.48\textwidth}
    \centering
    \includegraphics[width=\textwidth]{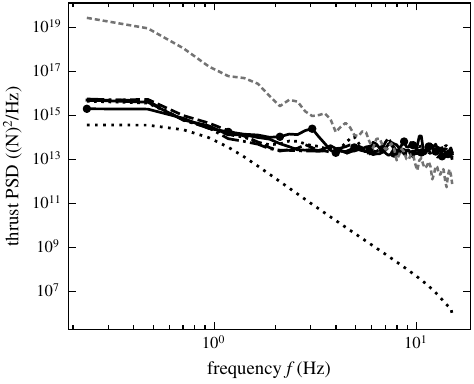}
    \caption{}\label{fig:psd_l4}
  \end{subfigure}
  \caption{Power spectral density of the collective-thrust command for policies trained at each fidelity level, tested under (a) L2 turbulence and (b) the L4 LES field at $U_\infty=8$\,m/s (crashed episodes excluded).}
  \label{fig:spectrum}  
\end{figure*} 
Figure~\ref{fig:spectrum} compares collective-thrust power spectra under L2 statistical turbulence and under the L4 LES field. Despite the markedly richer disturbance spectrum of the LES field, which carries up to two orders of magnitude more power across the 0.3--10\,Hz band (dashed curves), the control spectra of all five policies cluster within a factor of about two at every frequency, in both test winds. Fidelity-induced differences are therefore not frequency-selective: the policies do not fight different wind models in different bands. Complementing this picture, the thrust-saturation duty is identically zero for every learned policy in all 6750 evaluation episodes, including the 12\,m/s bins: none of the wind fields comes close to consuming the racer's control autority. PID is absent from the figure because none of its episodes survived these conditions long enough for spectral analysis.

 { The geometric controller, by contrast, fights the wind in a markedly narrower band: its thrust spectrum rolls off steeply beyond ${\sim}2$\,Hz in both test winds, an order of magnitude below the learned policies at 10\,Hz, consistent with a smooth PD-type law holding station against the disturbance rather than the RPM-level chatter of the learned policies.}

\subsection{{Trajectory-level illustration in the flow field}}

\begin{figure*}[!tbp]
  \centering
  \begin{subfigure}[t]{0.48\textwidth}
    \centering
    \includegraphics[width=\textwidth]{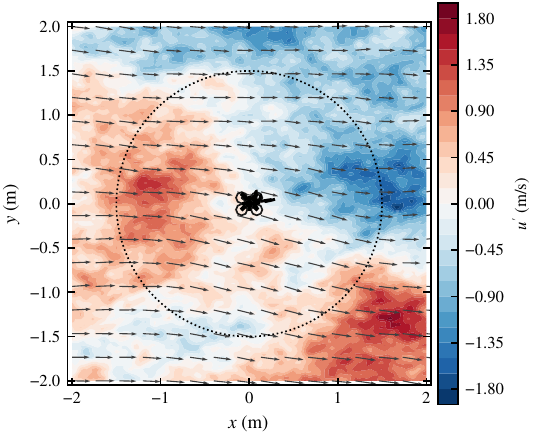}
    \caption{}\label{fig:compare_plan}
  \end{subfigure}\hfill
  \begin{subfigure}[t]{0.48\textwidth}
    \centering
    \includegraphics[width=\textwidth]{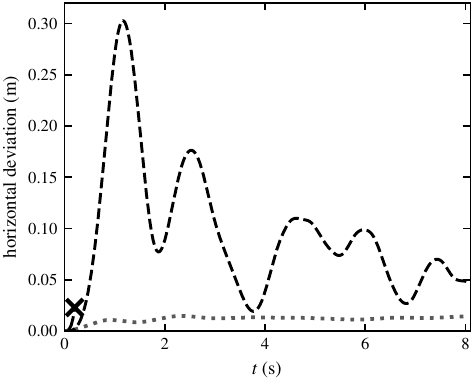}
    \caption{}\label{fig:compare_dev}
  \end{subfigure}
  \caption{The three controller families in one shared L3 wind realization: (a) plan view at hover altitude; (b) horizontal deviation from the target versus time.}
  \label{fig:compare}
\end{figure*}  
Figure~\ref{fig:compare} makes the three control laws tangible in a single shared L3 realization (L2 background, $\mathrm{TI}=10\,\%$, $U_\infty=8$\,m/s, plus two Oseen vortices; identical seed for all controllers). Figure~\ref{fig:compare}a is the background panel that shows the instantaneous streamwise fluctuation field ($u'$ at $t=4$\,s) at the hover altitude with the drone drawn to scale at the target. 

The PID baseline, flying transferred gains, departs the commanded attitude during the initial climb and is terminated at $t=0.2$\,s; the geometric controller pins the vehicle to within ${\sim}1$\,cm of the target for the full episode; and the L1-trained PPO policy holds station to within ${\sim}0.3$\,m with a characteristic slow oscillation. The panel also exposes the two vortices straddling the hover point---coherent structure that no stationary Gaussian field contains.

\begin{figure}[!tbp]
  \centering
  \includegraphics[width=3.5in]{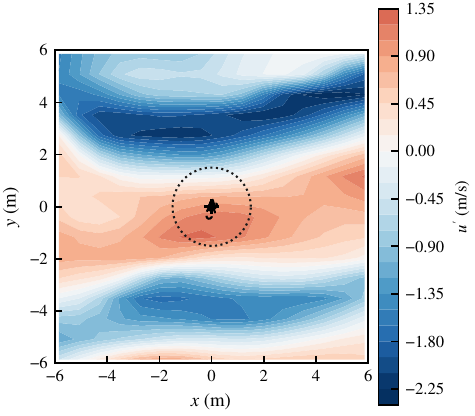}
  \caption{Plan view of the L4 LES disturbance field at hover altitude: streamwise fluctuation $u'$ with the L1-trained PPO flight path (dashed) and containment boundary (dotted).}
  \label{fig:flowplan}
\end{figure}
\begin{figure}[!tbp]
  \centering
  \includegraphics[width=3.5in]{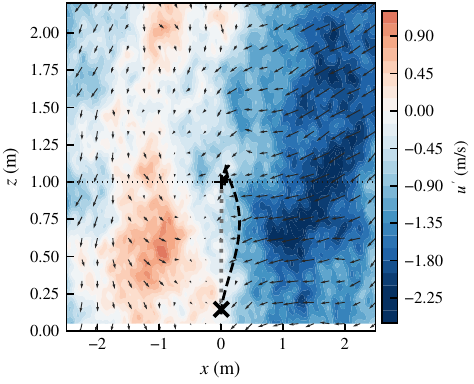}
  \caption{Side ($x$--$z$) view of the trajectories of Fig.~\ref{fig:compare} at hover altitude: $-.-$, PID; $\dots$, geometric SE(3); $--$, PPO; $\times$, PID termination.}
  \label{fig:flowside}
\end{figure}
{Figure~\ref{fig:flowplan} further complements this with a plan view of the L4 LES field at the hover altitude, overlaid with the trajectory flown by the L1-trained PPO policy in an actual L4 evaluation episode (the LES replay phase of that episode was recovered from the logged wind record). Figure~\ref{fig:flowside} gives the corresponding side ($x$--$z$) view through the hover point for the shared L3 realization of Fig.~\ref{fig:compare}: the vehicle rides the vertical gust structure, and the PPO policy's station-keeping excursions correlate visibly with the passage of the coherent upwash/downwash bands. 

\begin{figure}[!tbp]
  \centering
  \includegraphics[width=3.5in]{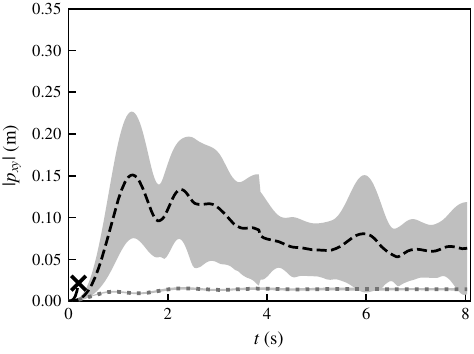}
  \caption{Mean horizontal deviation from the hover target versus time in the L3 wind of Fig.~\ref{fig:compare} (shaded: $\pm1$ std): $-.-$, PID; $\dots$, geometric SE(3); $--$, L1-trained PPO; $\times$, PID termination at $t\approx0.2$\,s.}
  \label{fig:controlwind}
\end{figure}
Finally, Fig.~\ref{fig:controlwind} aggregates the control performance of the three families in wind: mean horizontal deviation from the target versus time, with $\pm1$ standard-deviation bands. Two features stand out: every transferred-gain PID episode terminates at $t\approx0.2$\,s---a takeoff-phase departure, present even at 4\,m/s, rather than a gust-specific failure---and the geometric controller's error band is an order of magnitude tighter than the learned policy's, whose ${\sim}0.1$\,m mean excursion with its slow oscillation is the visible price of RPM-level chatter.

\subsection{Platform sensitivity}

\label{sec:platform}
A companion campaign under the identical protocol on the 27\,g cf2x model (five levels, five seeds, 5040 episodes) paints a quantitatively different but structurally consistent picture. There, training-fidelity effects were larger---crash rates under 8\,m/s L2 turbulence ranged from 36\,\% (L0-trained) to 58\,\% (L3-trained), and the 50\,\% crash boundary lay between 7.4 and 8.9\,m/s for the learned policies and 6.5\,m/s for PID---yet still second-order relative to the platform's physical envelope, and the same L1-best ranking held. 

{The geometric controller reproduces the same pattern with classical means: on cf2x it trails the natively tuned PID in calm and light wind (no integral action, hence a steady-state offset under sustained drag), but beyond 8\,m/s it overtakes both PID and most learned policies---at $U_\infty=8$\,m/s L2 turbulence its crash rate is 0--10\,\% against 60--90\,\% for PID, with correspondingly lower RMSE---until the 12\,m/s bin exceeds the platform's physical speed envelope and all controllers fail alike.} 

The governing dimensionless quantity is the wind-induced acceleration relative to $g$: wind enters through drag, and at 8\,m/s the disturbance acceleration is of order $8\,\mathrm{m/s^2}$ on the 27\,g cf2x but only $0.5\,\mathrm{m/s^2}$ on the 0.83\,kg racer (equal drag coefficient, thirty times the mass, comparable rotor-speed sums). Flow fidelity therefore matters in proportion to platform wind sensitivity: the flatter the disturbance-to-control-authority ratio, the flatter the cross-fidelity matrix.

\section{Discussion}\label{sec:discussion}

\subsection{Practical guidance}

The above results answer the initial question directly, and the answer splits by platform. For the racing-class vehicle studied here, wind-field{fidelity} is at most a third-order design variable within the swept envelope: every domain-randomized training level (L1--L4) yields policies that survive the full 0--12\,m/s range at $\le 14\,\%$ crash rate, and even no-wind training degrades gracefully. If one training wind must be chosen, the evidence favors the cheapest structured one: discrete-gust (1-cosine) domain randomization ranks first in every test column of the racer matrix and held the same position in the cf2x campaign---thirty train--test cells across two platforms with a consistent ordering. 

The expensive options (coherent-structure injection, LES replay) buy nothing measurable on top of it for robustness; their value lies elsewhere, e.g.\ in high-bandwidth disturbance-rejection benchmarking or in estimator-in-the-loop studies where the {information content} of the wind field matters. For wind-sensitive micro platforms (cf2x-scale), the same ranking applies but the absolute margins are compressed against the physical speed envelope; there the first-order investment is platform authority, not wind realism. 

{The geometric-baseline result adds one qualifier for practice: on both platforms, a fixed-gain SE(3) controller tuned on the deployment platform matched or beat every learned policy in steady-state accuracy at high wind---so where a model of the platform is available, classical geometric control remains the stronger default, and the case for learned controllers rests on the settings where model quality or adaptivity, not gust structure, is the binding constraint.}

\subsection{{Toward real flight}}

The natural next question is how these simulation results should be carried to a real vehicle---and whether a policy trained in this framework can be deployed directly, zero-shot. 

The present evidence supports a qualified yes for high-authority platforms, on two conditions. First, the sim-to-real gap that remains after this study is not in the wind field: the cross-fidelity matrix shows that structural fidelity of the training disturbance contributes at most third-order robustness differences, so the residual gap lives in the {vehicle} model---motor and ESC dynamics, propeller thrust curves at oblique inflow, IMU noise and state-estimation latency---which the zero-shot ablations of Chen et al.~\cite{chen2024whatmatters} likewise identify as decisive. A deployment pipeline should therefore spend its identification budget on actuator and sensor models, not on more realistic turbulence. 

Second, domain randomization must cover the deployment envelope: the flat matrix holds because training sampled the full $0$--$12$\,m/s, 5--20\,\% TI grid, and the cheapest structured generator (L1 gusts) already suffices. Under these two conditions, a first zero-shot flight of a gust-randomized PPO policy on a racer-class platform is defensible, ideally staged through motion-capture hover, fan-array disturbance tests, and only then outdoor operation.

{For wind-sensitive micro platforms the answer changes: margins compress against the physical speed envelope (Sec.~\ref{sec:platform}), and zero-shot transfer of a fixed policy is no longer the robust default. There the evidence points to a layered architecture---a geometric SE(3) tracking law (with an online disturbance estimate added to Eq.~\eqref{eq:geomforce}) as the certifiable inner loop, a learned policy for the outer loop trained under gust randomization, and an onboard wind estimate~\cite{tasim2026windaware,kiehn2024empiricalbayes,gahan2025hinftywind} feeding both. Residual fine-tuning on flight data, rather than from-scratch relearning, closes the remaining gap. In both cases the flight-readiness question is decided by platform authority and model identification, not by the structural fidelity of the training wind.}

\subsection{Why the PID approach fails}

The PID baseline's 100\,\% crash rate in wind on the racer is primarily a gain-transfer artifact: the shipped differential-flatness gains were tuned for a 27\,g Crazyflie and, by the fixed-gain principle of Sec.~\ref{sec:control}, were not re-tuned for a platform thirty times heavier---only the physically determined quantities (mass, thrust and torque constants, etc.) were adapted. 

{The geometric SE(3) row is the control experiment that isolates this artifact: subject to the same fixed-gain discipline but tuned once on the racer's own L0 task, it survives every windy column at 0\,\% crash rate with centimetre-level steady-state error. The PID failure is therefore attributable to the transferred gains, not to classical control per se---and, symmetrically, the learned policies' advantage over PID on this platform should not be read as an advantage over well-tuned classical control.} 

The failure mode---{attitude-limit departure during the climb transient under gust loading}---is nevertheless {consistent with} the mode that limited the (natively tuned) PID row in the cf2x campaign at high wind, where it crossed 50\,\% crash rate well inside the learned policies' envelope. We therefore report the row as a conservative fixed-gain reference: a platform-tuned PID would do better, but the qualitative contrast---fixed-gain controllers trade position accuracy for survival far less gracefully than learned policies under sustained turbulence---stands in both campaigns.

\subsection{Limitations}

We wish to emphasize that currently the study is simulation-only: no flight-test validation is performed, so conclusions about sim-to-real deployment are inferential, resting on the similarity of the simulated relative-wind chain to its physical counterpart. {The rigor established in Sec.~\ref{sec:datarigor} is accordingly that of a validated simulation chain---standards-anchored generators, a log-law-validated LES, and a peer-reviewed vehicle model---and not of flight data; within the simulation, statistical resolution is bounded by the five-seed protocol, so effects smaller than the seed-level spread are reported as such rather than claimed.} Rotor and body aerodynamics are represented by the vehicle simulator's models rather than by blade-resolved computation, and structural flexibility of the airframe is neglected; aeroelastic coupling could move energy into frequency bands this study treats as disturbance-free. The L4 ensemble is finite---a single boundary-layer realization, velocity-rescaled for non-native wind conditions---so rare extremes of the boundary layer may be under-sampled relative to what a long deployment would encounter. The two campaigns differ in training budget ($3\times10^6$ steps on cf2x versus $10^6$ on the racer), though all reported policies are converged by their held-out evaluation curves. Finally, the controller panel spans the fixed-gain and offline-learned ends of the design spectrum; online-adaptive designs (e.g.\ model-reference adaptive control~\cite{lavretsky2013robust}), whose adaptation rate interacts with gust transients in a fundamentally different way, are left to future work within the same protocol.

\section{Conclusions}\label{sec:conclusions}

This paper set out to answer a deliberately narrow question with broad practical consequences: does the physical fidelity of the training wind field matter for the robustness and transferability of learned quadrotor controllers? To address it, we defined a five-level hierarchy of wind-disturbance fidelity (L0--L4), spanning wind-free flight, discrete deterministic gusts, stationary statistical turbulence, synthetic coherent structures, and LES-derived flow fields, with dimensionless matching (turbulence intensity, integral-scale ratio, spectral shape) enforced across levels. We executed a full train-fidelity $\times$ test-fidelity cross-evaluation protocol on a racing-class quadrotor in hover, with PPO policies trained at each level and evaluated at every other, alongside two training-free classical baselines: a cascaded PID controller and a geometric SE(3) tracking controller. A companion campaign on a wind-sensitive 27~g platform provided the platform-sensitivity contrast.

The answer is clear and platform-dependent. On the racing-class platform, the cross-fidelity matrix is remarkably flat: position RMSE varies only between $0.29$ and $0.33$~m and crash rates between $0$ and $14\,\%$ across all twenty train--test pairs at 8~m/s, with no diagonal advantage evident. More strikingly, the cheapest structured training wind---discrete-gust domain randomization (L1)---ranks first in every test column, a ranking replicated on the cf2x platform. The expensive options (L3, L4) buy no measurable robustness advantage on top of it. Control-input spectra confirm the mechanism: despite the LES field carrying up to two orders of magnitude more disturbance power than the statistical turbulence, the learned policies respond with fidelity-independent spectral economy and zero thrust saturation throughout.

The geometric SE(3) baseline qualifies the classical-control side of the story. PID with gains transferred from the 27~g platform fails outright in wind (100\,\% crash rate at 8~m/s), but the same fixed-gain discipline applied with platform-tuned geometric gains survives the entire envelope at zero crash rate and with the tightest station-keeping of any controller tested---centimetre-level steady-state error against the learned policies' decimetre-level excursions. The PID failure is therefore a gain-transfer artifact, not a classical-control ceiling; the learned policies' advantage over the transferred-gain baseline should not be misread as an advantage over well-tuned classical control.

Three practical conclusions follow for the design of learning-based flight control systems. First, for high-authority platforms (thrust-to-weight $\gtrsim 4$), wind-field fidelity is at most a third-order design variable within the certification-relevant envelope; domain randomization over cheap deterministic gusts suffices, and LES-grade flow fields are unnecessary for robustness. Second, for wind-sensitive micro platforms, the same ranking applies but the absolute margins compress against the physical speed envelope; there the first-order investment is platform authority and online wind estimation, not wind realism in the training environment. Third, where a platform model is available, classical geometric control with platform-tuned gains remains the stronger default, matching or outperforming learned policies in steady-state accuracy; the case for learned controllers rests on settings where model quality, adaptation rate, or unstructured task complexity---not gust structure---is the binding constraint.

Several limitations of the present study point directly to future work. The results are simulation-only; no flight-test validation is performed, so conclusions about sim-to-real deployment are inferential. The vehicle aerodynamics are modeled at the level of the simulator's relative-velocity drag law rather than blade-resolved computation, and aeroelastic coupling is neglected. The LES ensemble is finite---a single boundary-layer realization, velocity-rescaled for non-native wind conditions---so rare extreme events are under-sampled. Online-adaptive designs (model-reference adaptive control, adaptive disturbance observers) are not included in the controller panel and may interact with gust transients in ways that this fixed-gain versus offline-learned comparison does not capture.

These limitations define the next steps. Flight-test validation of the L1-trained policy on a racer-class platform, staged through motion-capture hover and fan-array disturbance tests, would establish whether the flat-matrix result survives the sim-to-real gap. Extension to trajectory tracking rather than hover, and to aggressive maneuvering where the aerodynamic coupling between vehicle motion and the wind field is stronger, would test the generality of the fidelity-ranking result. Incorporation of online-adaptive designs (e.g., disturbance-observer-based control, model-reference adaptive control) into the same cross-fidelity protocol would reveal whether adaptation rate, rather than training-fidelity, is the dominant robustness factor. Finally, the LES production workflow established here---a single-GPU Boussinesq solver validated against the log law---is now available for reuse in future studies that do require high-fidelity flow fields, such as estimator-in-the-loop experiments where the information content, not just the energy content, of the disturbance field is the independent variable.

In summary, this paper establishes that for high-authority quadrotor platforms, the structural fidelity of the training wind can be traded away against magnitude randomization with no measurable robustness penalty. The cheapest structured generator---discrete-gust domain randomization---ranks first across thirty train--test cells on two platforms. Wind-fidelity investment should therefore scale with platform wind sensitivity, and the marginal value of LES-grade flow fields for robustness training on this class of vehicles is, within the swept envelope, zero. {All of these conclusions rest on the validated data chain of Sec.~\ref{sec:datarigor}: every disturbance generator is checked quantitatively against its analytical or certification-standard reference, the LES fields against the imposed log law, and the vehicle dynamics are taken from a peer-reviewed, system-identified benchmark.}

\section*{Acknowledgements}
This work is partly supported by National Science Foundation of China (Grant No. 12432016 \& No. 12272007)  and Shĳiazhuang-Peking University Cooperation Special Program. The author has utilized Kimi to prepare the simulation running and sketching scripts. The code will be released on GitHub upon acceptance of this paper.


\end{document}